\documentclass{sig-alternate-2013}

\usepackage[T1]{fontenc}
\usepackage{times}

\usepackage[
pass,
]{geometry}

\toappear{To appear in MSM 2017: 8th International Workshop on Modelling Social Media: Machine Learning and AI for Modelling and Analysing Social Media, 2017 International World Wide Web Conference (\emph{WWW`17}) Workshop, April 2017, Perth, Australia.}

\permission{\copyright 2017 International World Wide Web Conference Committee (IW3C2), 
published under Creative Commons CC BY 4.0 License.}
\conferenceinfo{WWW 2017,}{April 3--7, 2017, Perth, Australia.}
\copyrightetc{ACM \the\acmcopyr}
\crdata{978-1-4503-4914-7/17/04. \\
http://dx.doi.org/10.1145/3041021.3053903 \\
\includegraphics{cc.png}
}

\begin{document}

\title{The nature and origin of heavy tails in retweet activity}
%
%
%
%
%

\numberofauthors{4} 
%
\author{
%
%
\alignauthor
Peter Mathews\\\
\affaddr{The University of Adelaide}\\
\affaddr{Adelaide, Australia}\\
\email{\normalsize peter.mathews@adelaide.edu.au}\\
\alignauthor Lewis Mitchell\\
\affaddr{The University of Adelaide}\\
\affaddr{Adelaide, Australia}\\
\email{\normalsize lewis.mitchell@adelaide.edu.au}\\
\and
\alignauthor Giang T. Nguyen\\
\affaddr{The University of Adelaide}\\
\affaddr{Adelaide, Australia}\\
\email{\normalsize giang.nguyen@adelaide.edu.au}\\
\alignauthor Nigel G. Bean\\
\affaddr{The University of Adelaide}\\
\affaddr{Adelaide, Australia}\\
\email{\normalsize nigel.bean@adelaide.edu.au}
}



\maketitle
\begin{abstract}
Modern social media platforms facilitate the rapid spread of information online. Modelling phenomena such as social contagion and information diffusion are contingent upon a detailed understanding of the information-sharing processes. In Twitter, an important aspect of this occurs with retweets, where users rebroadcast the tweets of other users. To improve our understanding of how these distributions arise, we analyse the distribution of retweet times. We show that a power law with exponential cutoff provides a better fit than the power laws previously suggested. We explain this fit through the burstiness of human behaviour and the priorities individuals place on different tasks.
\end{abstract}


\section{Introduction}
Twitter is one of the most popular social media sites with twitter.com being the 10th most visited website in the world \cite{alexa:website}. The main method of interaction on Twitter is by users changing their status, known as a \emph{tweet}. Other users can interact with this tweet in several ways including favouriting or retweeting the tweet. The temporal dynamics of how information spreads through Twitter through retweets provides an excellent case study for understanding information propagation.

There exists a large body of work on cascades in social media systems, e.g. \cite{Kupavskii:2012:PRC:2396761.2398634, Wu:2011:SWT:1963405.1963504, bakshy2011identifying, Li:2014:MBT:2566267.2566312, Lu2014747}. These focus on the size and volume aspect of a cascade, often using statistics or machine learning techniques to predict the final cascade size based on various features. However, the temporal component underlying this phenomena is relatively poorly understood. There also exists generative behavioural models for human dynamics and large scale collective phenomena using stochastic models, e.g. \cite{barabasi05}. The link between the two may not always be clear.

This work builds upon both research topics, using social behavioural models to explain the temporal component of information cascades in a social media system.

We make the following key new contributions:
\begin{itemize}
	\item Showing that a power law with exponential cutoff is a better fit for the distribution of retweet times than a power law.
	\item Providing an explanation of the origin of the power law with exponential cutoff for the retweet time distribution.
\end{itemize}

The remainder of the paper is structured as follows: In Section \ref{sec:priorwork} we review prior work on Twitter dynamics and causes of power laws. In Section \ref{sec:retweettimeanalysis} we introduce the dataset and analyse the best way to fit a distribution. In Section \ref{sec:originofpowerlaws} we explain the underlying processes which lead to the distribution. In Section \ref{sec:conclusions} we summarise our findings and discuss possible extensions to this work.

\section{Related work}
\label{sec:priorwork}

\subsection{Twitter dynamics}

There exists a significant amount of related work about modelling Twitter dynamics. However, although other authors have touched on the subject, the distribution of retweet times has not been analysed in detail previously.

Crane and Sornette \cite{Crane:2008p6660} studied the response of a social system after endogenous and exogenous bursts of activity. They found that after the initial peak, activity declines as a power law distribution.

Zhao \emph{et al.} \cite{Zhao:2015:SSP:2783258.2783401} looked at the reaction time for retweets from an initial tweet. They plotted the retweet times up to 15 hours after the initial tweet and concluded that the linear trend on logarithmic axes suggests a power law decay. 

Lu \emph{et al.} \cite{Lu2014747} developed a method to model the lifetime number of retweets from an originating source. They found the distribution to be a power law with exponent in the range 0.6 to 0.7. They proposed that the ``probability of being forwarded is proportional to the product of preferential attachment and transmissibility". Wu \emph{et al.} \cite{Wu:2011:SWT:1963405.1963504} performed an extensive analysis of the production, flow and consumption of information on Twitter. They found that different content types exhibit dramatically different characteristic lifespans.

There has also been much work in predicting cascade size in social media structures based on various factors. Kupavskii \emph{et al.} \cite{Kupavskii:2012:PRC:2396761.2398634} predicted the size of the cascade based on the initial spread using machine learning techniques. Bakshy \emph{et al.} \cite{bakshy2011identifying} looked at the possibility to achieve cascades through social media structures from ordinary influencers.

Bild \emph{et al.} \cite{Bild:2015:ACU:2745838.2700060} showed that lifetime tweet counts follow a type-II discrete Weibull distribution. They showed that the tweet rate distribution is asymptotically power law but exhibits a lognormal cutoff over finite sample intervals. They also showed that the intertweet interval distribution for a single user is power law with exponential cutoff.

Doerr \emph{et al.} \cite{10.1371/journal.pone.0064349} showed that many processes governing online information spread have a log-normal distribution. The authors questioned the applicability of fitting power law distributions to temporal behavioural data. They argued that the low exponents found in temporal data militates against preferential attachment. They also argued that while preferential attachment provides an explanation for scale-free degree distribution, it does not provide insight into propagation time distributions. Based on this, they claimed that there does not exist a theoretical model able to explain the observed traces of online human behaviour.

A clear shortcoming of this paper is that it only considered preferential attachment as the cause of power laws. Although preferential attachment is a common and well-known mechanism for the generation of power laws, it is certainly not the only mechanism.

User interest in topics has a tendency to decay exponentially over time \cite{Ding:2005:TWC:1099554.1099689, Li:2014:MBT:2566267.2566312}. This will form a component of our model in Section \ref{sec:originofpowerlaws} where we consider the user interest in tweets after a period of time.

\subsection{Causes of power law}
\label{sec:Causes}

Power laws occur frequently in nature and man-made systems. Examples of phenomena that can be modelled well by power laws include frequencies of words in most languages, sizes of earthquakes, intensity of wars, severity of terrorist attacks, sightings of bird species and many others \cite{Clauset:2009:PDE:1655787.1655789}.

Mitzenmacher \cite{Mitzenmacher_abrief} and Newman \cite{Newman05powerlaws} identified 14 causes of power laws, both natural and man made. In particular we note:
\begin{itemize}
	\item Growth by preferential attachment, where new entities attach to existing entities proportional to their current size \cite{1999PhyA..272..173B}.
	\item The inter-event time distribution for a single event type where behaviour is a consequence of a decision-based queuing process \cite{PhysRevE.73.036127}.
\end{itemize}

As these two causes of power law are the most relevant to our work we discuss them in more detail.

\subsubsection*{Growth by preferential attachment}

In preferential attachment, new entities attach to existing entities proportional to their current size. In Polya's Urn model \cite{Mahmoud:2008:PUM:1408784} where balls are added to urns with probability proportional to the number of balls in the urn, it can be shown that the number of balls per urn is distributed as a power law. Power laws by preferential attachment occur frequently in nature and in human sciences. Cities tend to grow proportional to their current size \cite{Ioannides2003127}. Networks have a tendency to grow by attaching new nodes to nodes that already have a large number of connections \cite{Barabasi99emergenceScaling}. 

\subsubsection*{Power law due to decision-based queuing process}

Barabasi \cite{barabasi05} showed that the bursty nature of human behaviour can be explained by a decision-based queuing process, which was further explained by Vazquez \emph{et al.} \cite{PhysRevE.73.036127}. Consecutive actions from a single user, such as the inter-event times between emails sent, have a tendency to be power law distributed. This is different to the exponential distribution that would occur if human activity was modelled as a Poisson process. Barabasi showed that the timings of five human activity patterns, email and letter based communications, web browsing, library visits and stock trading, followed non-Poisson statistics. When humans execute tasks based on some perceived priority, the waiting time between tasks is heavy-tailed.

\section{Retweet Time Analysis}
\label{sec:retweettimeanalysis}

\subsection{Overview of retweet rates}

We define the \emph{retweet rate} as the number of retweets per unit time occuring for a particular seed tweet. A tweet tends to have the highest retweet rate shortly after it is posted, with the retweet rate slowly decaying over time. We consider the distribution of times until retweets occur and look to determine the most appropriate model to represent this distribution.

Analysing retweet rate decay gives an insight into the longevity of interest in topics being tweeted. Retweets indicate interest about the tweet by a user, so a seed tweet with a slow retweet decay rate suggests that the topic of the tweet has longevity.

To illustrate the problem, we first look at specific examples of retweet time distributions.

\begin{figure*}[ht]
	\begin{center}
		\begin{tabular}{@{}c@{}c@{}c@{}c}
			\includegraphics[width=2in]{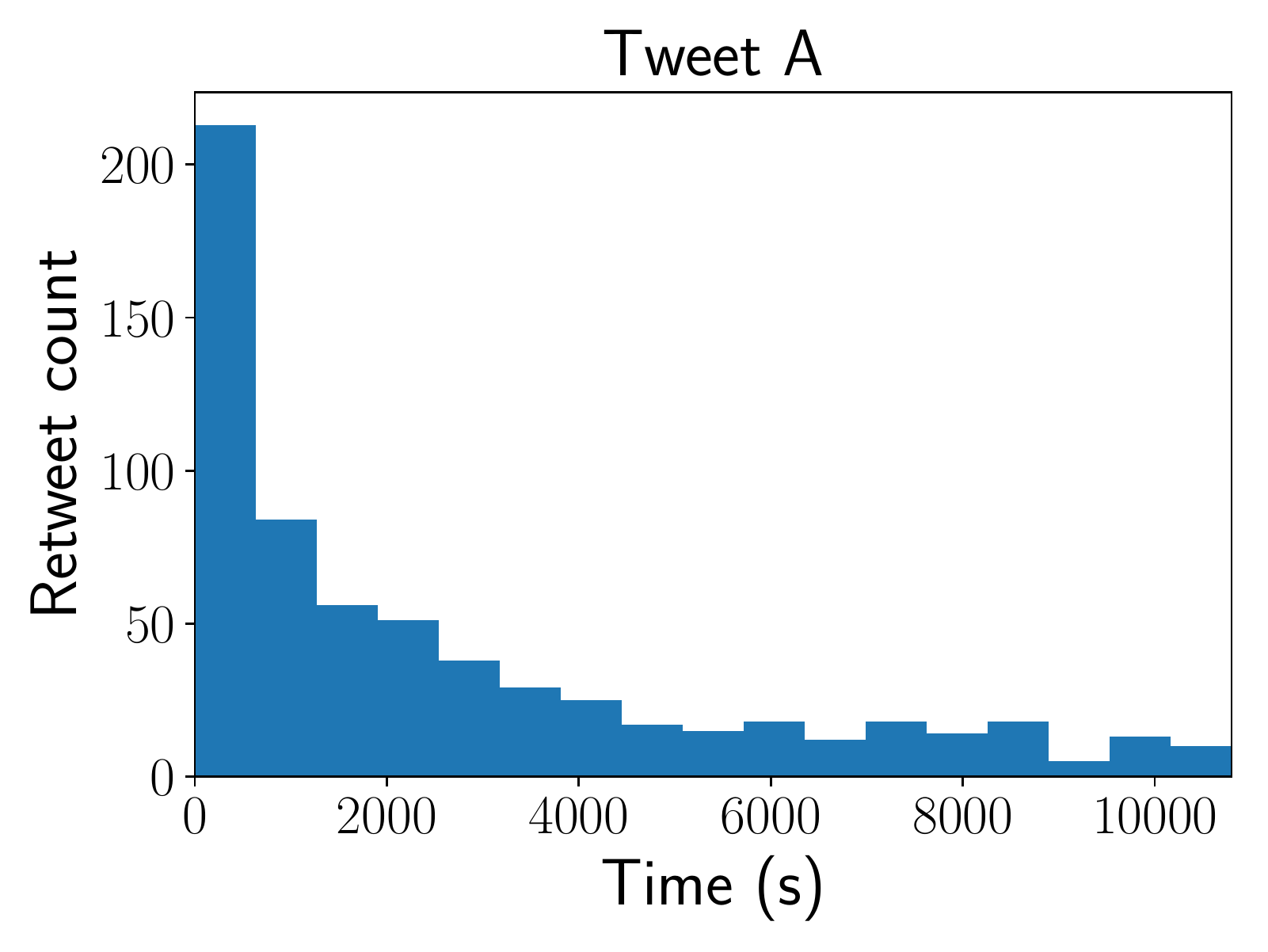} \ &
			\includegraphics[width=2in]{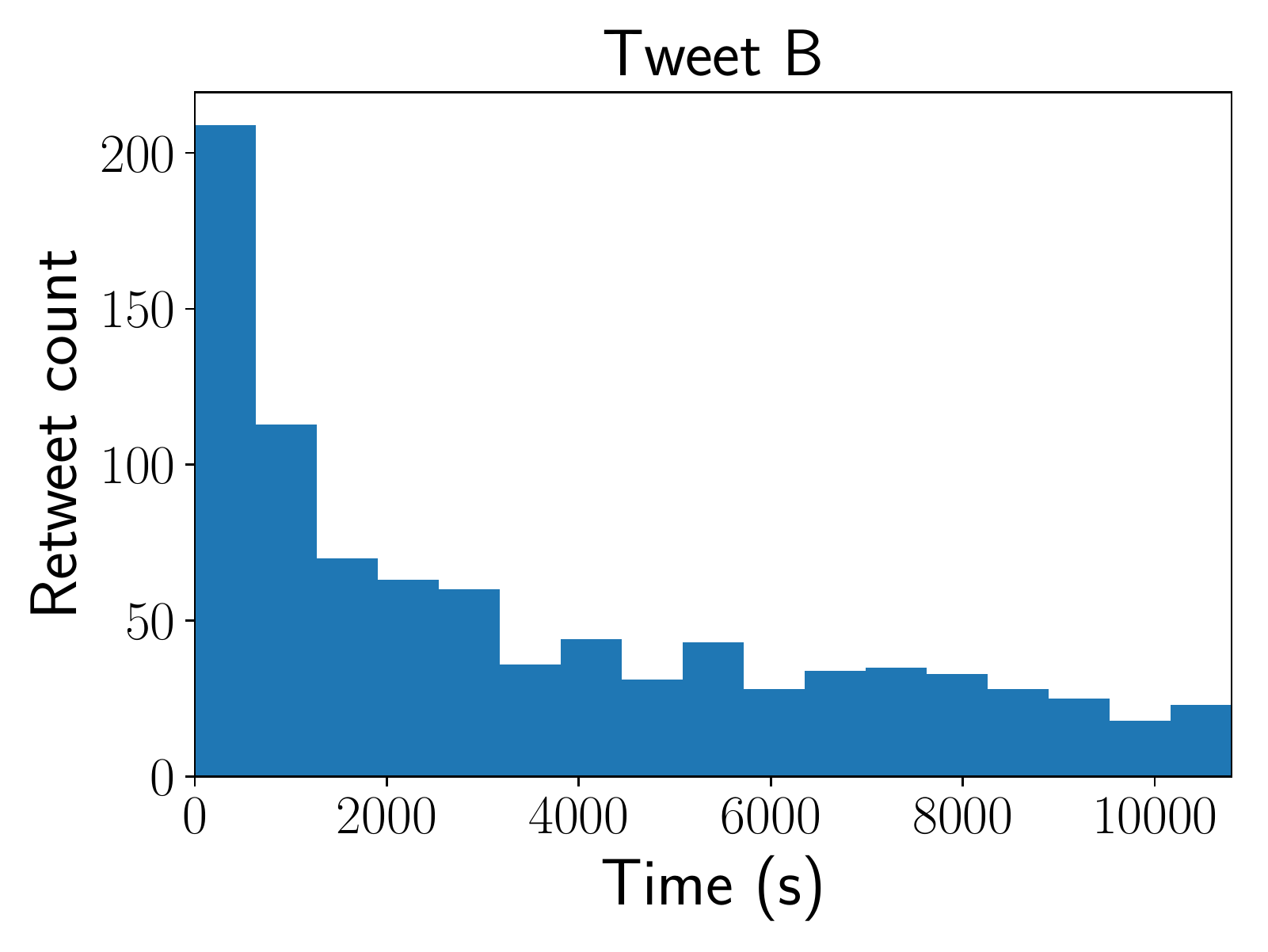} \ &
			\includegraphics[width=2in]{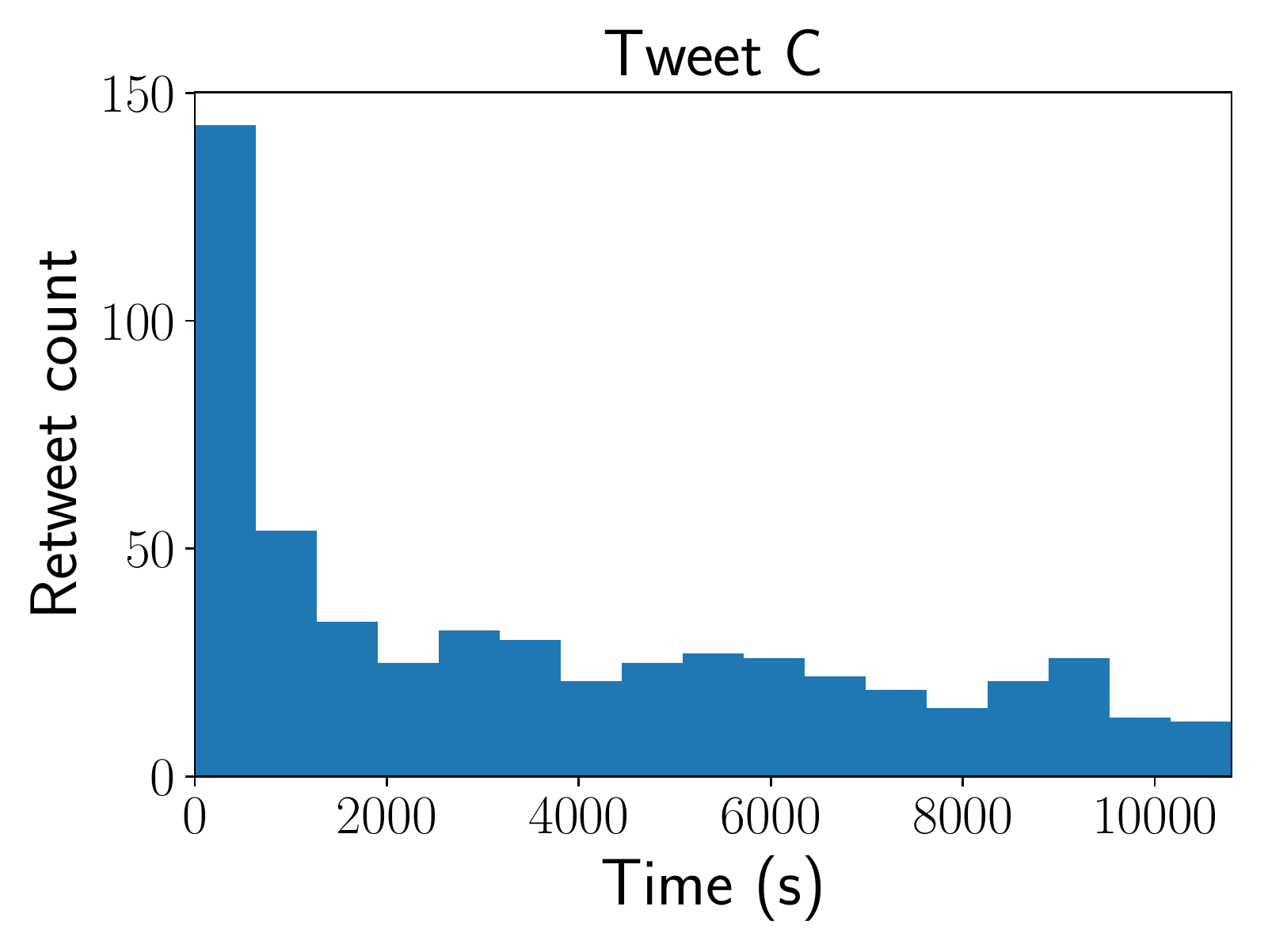} \ \\		
			\includegraphics[width=2in]{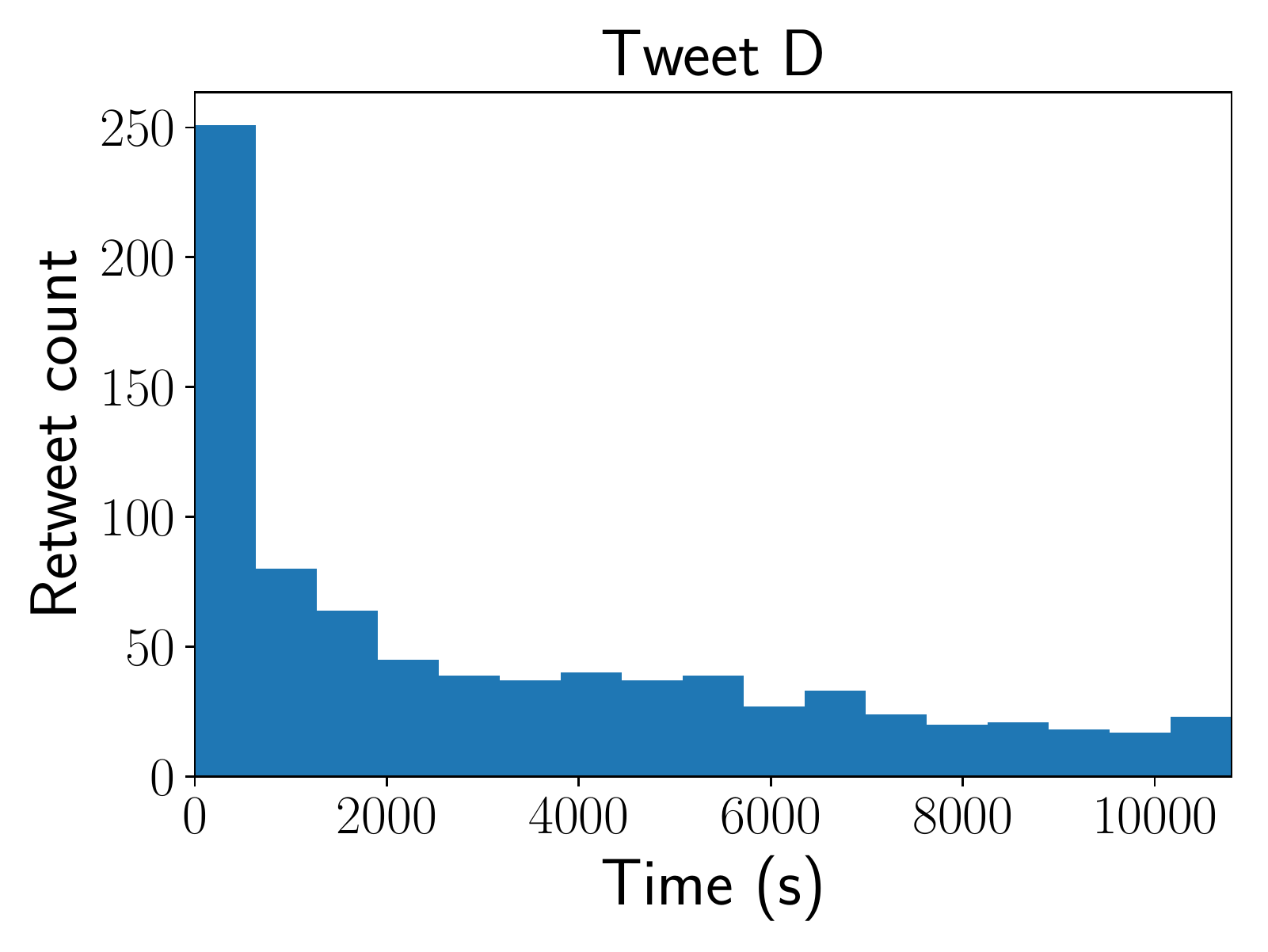} \ &
			\includegraphics[width=2in]{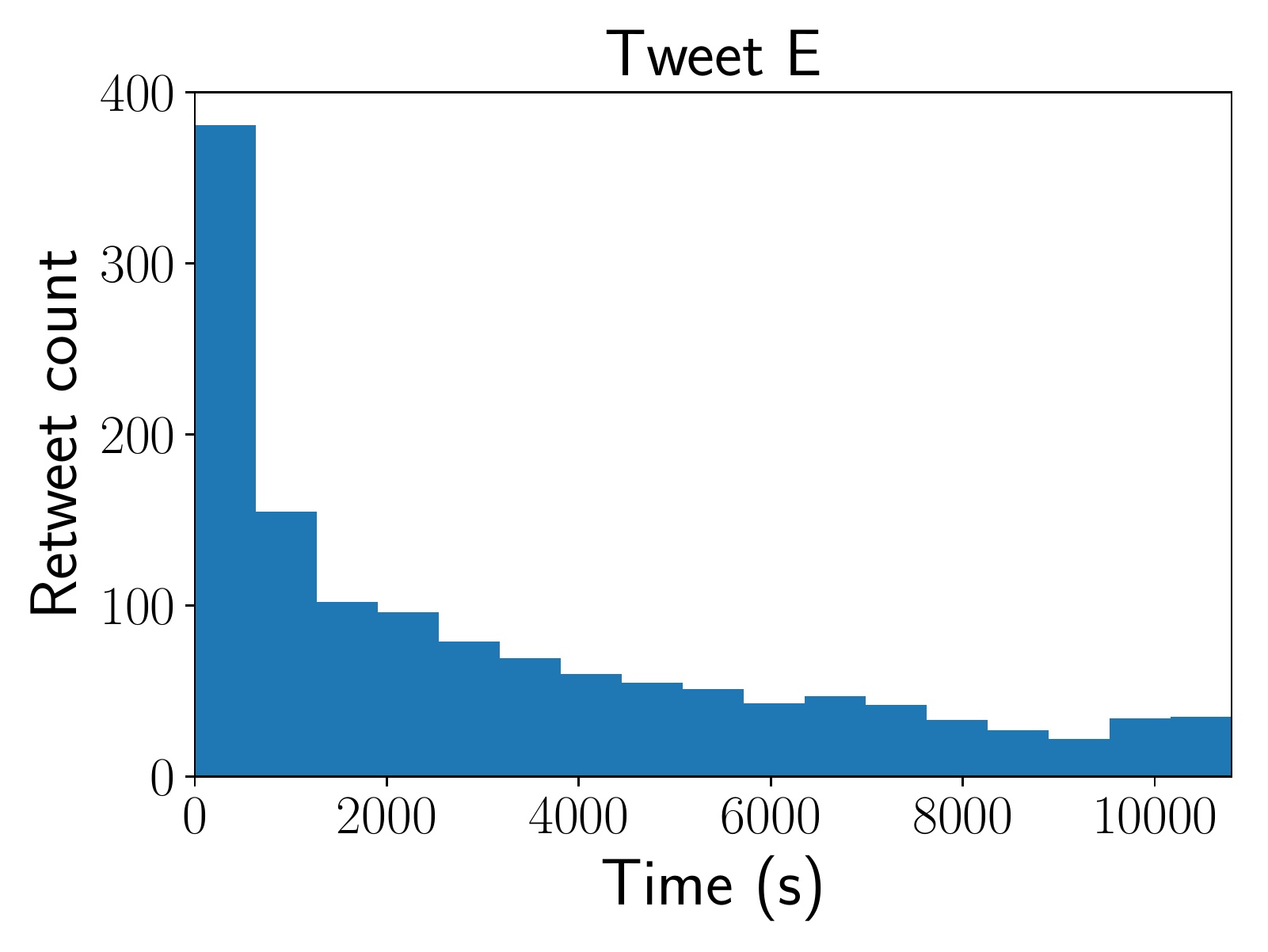} \ &		
			\includegraphics[width=2in]{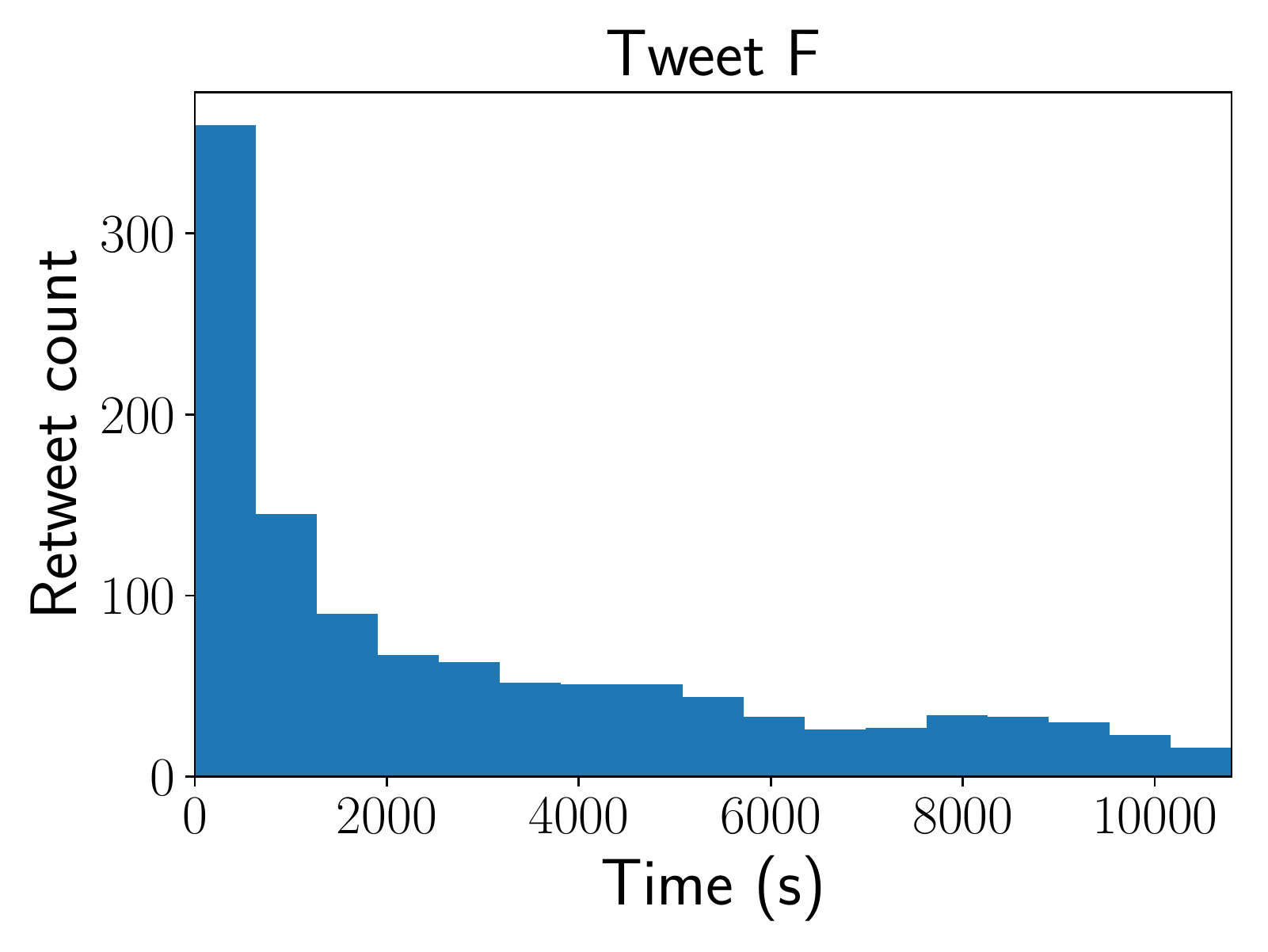} \ \\\
		\end{tabular}
	\end{center}
	\caption{Retweet count histograms showing the first three hours after the initial tweet. The rate of retweets tends to decay over time.}
	\label{fig:RetweetDecayHist}
\end{figure*}

Figure \ref{fig:RetweetDecayHist} shows examples of retweet counts with constant-width bins from six sample seed tweets by Donald Trump (Twitter: @realDonaldTrump) in February 2016. As can be seen, the retweet counts decay from their starting levels with some amount of noise.

\begin{figure*}[!h]
	\begin{center}
		\begin{tabular}{@{}c@{}c@{}c@{}c}
			\includegraphics[width=2in]{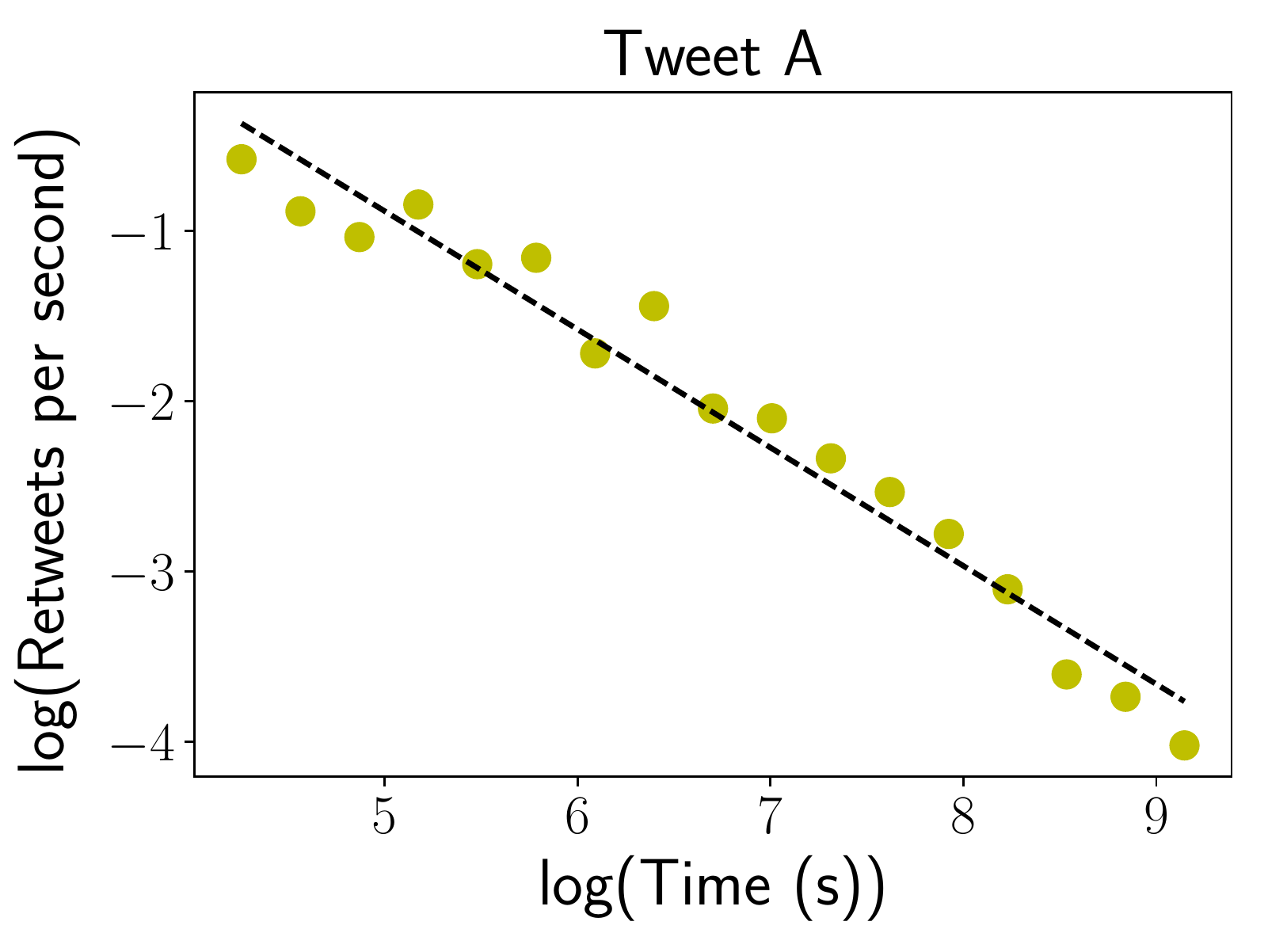} \ &
			\includegraphics[width=2in]{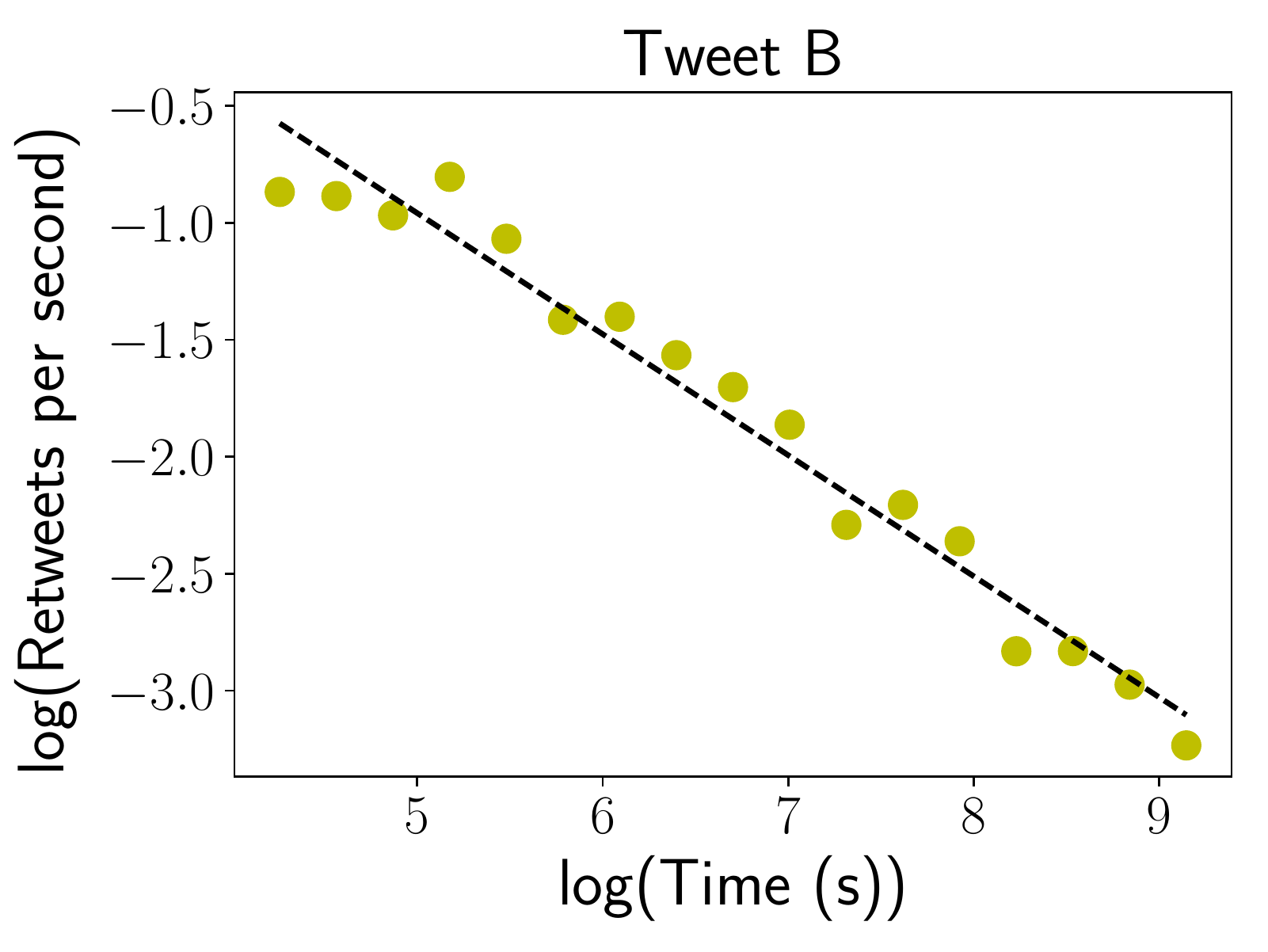} \ &
			\includegraphics[width=2in]{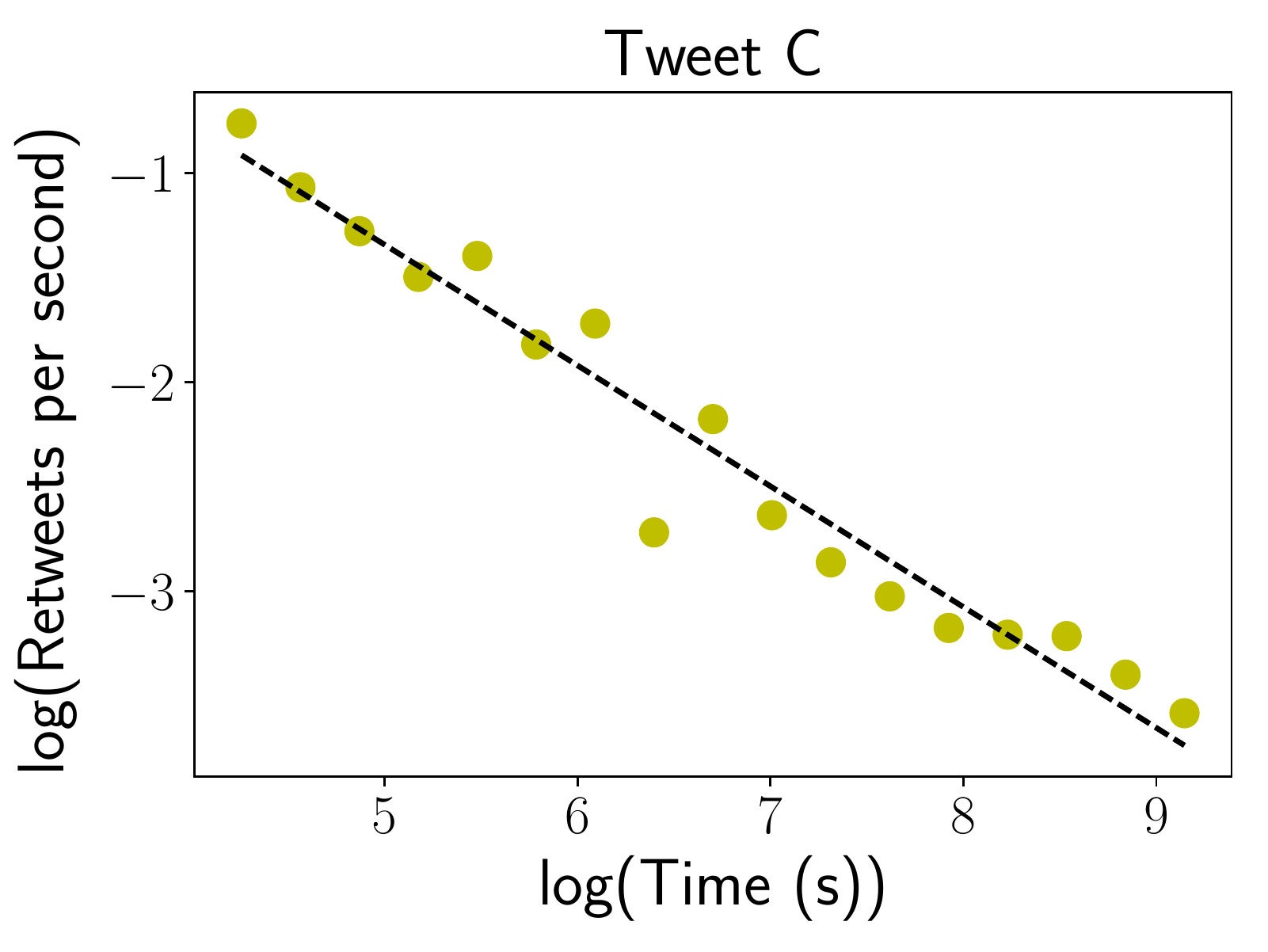} \ \\		
			\includegraphics[width=2in]{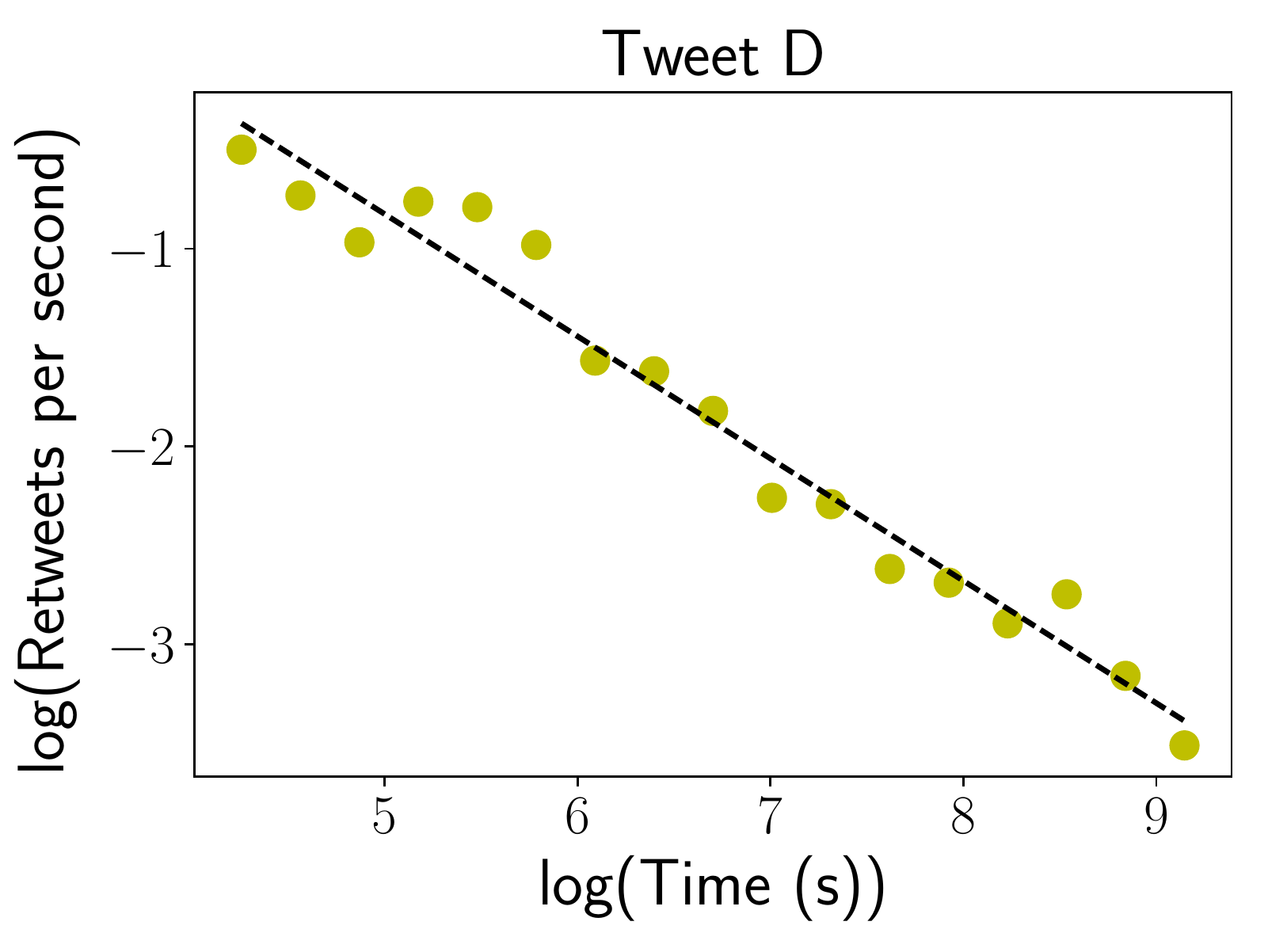} \ &
			\includegraphics[width=2in]{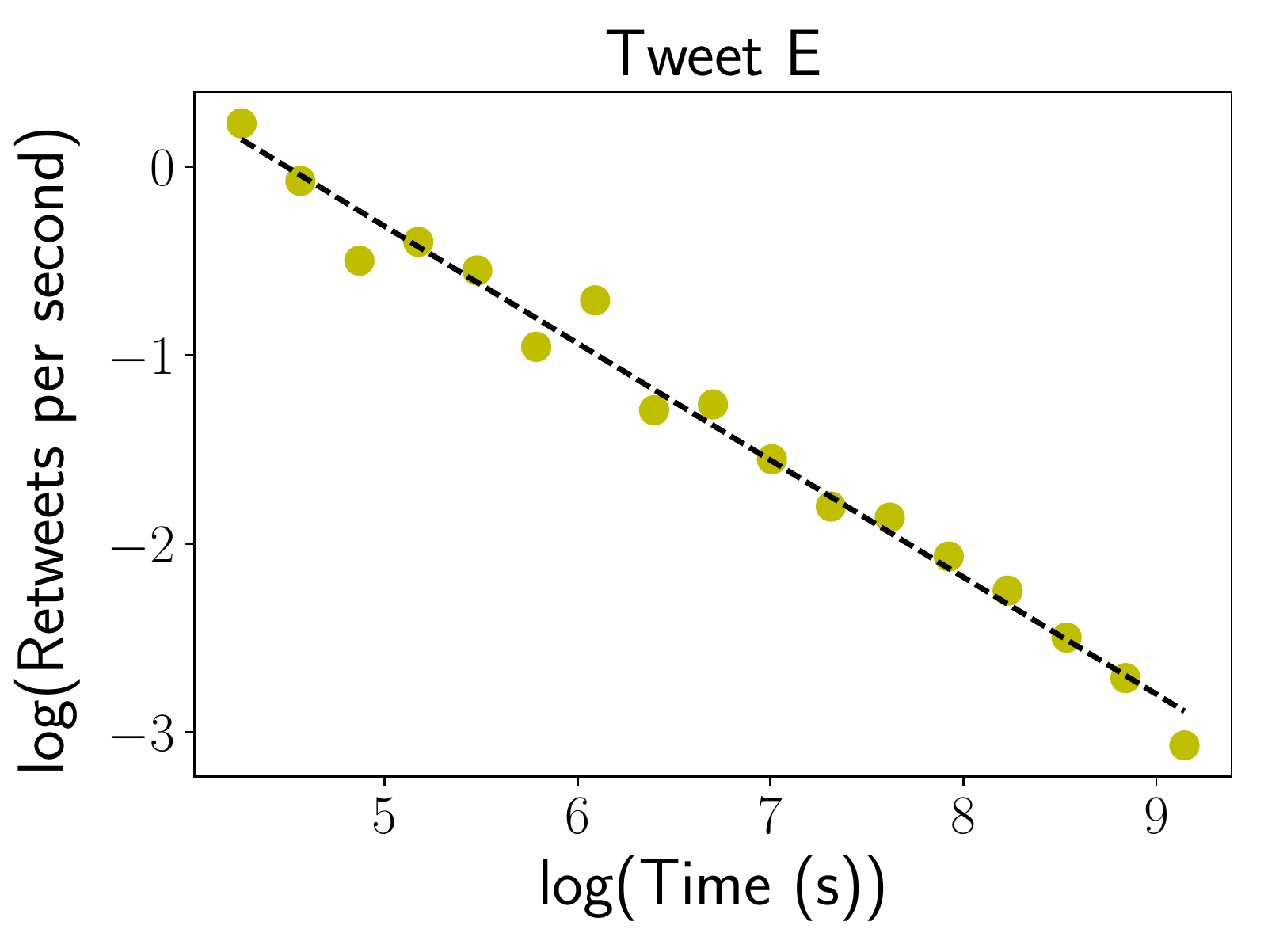} \ &		
			\includegraphics[width=2in]{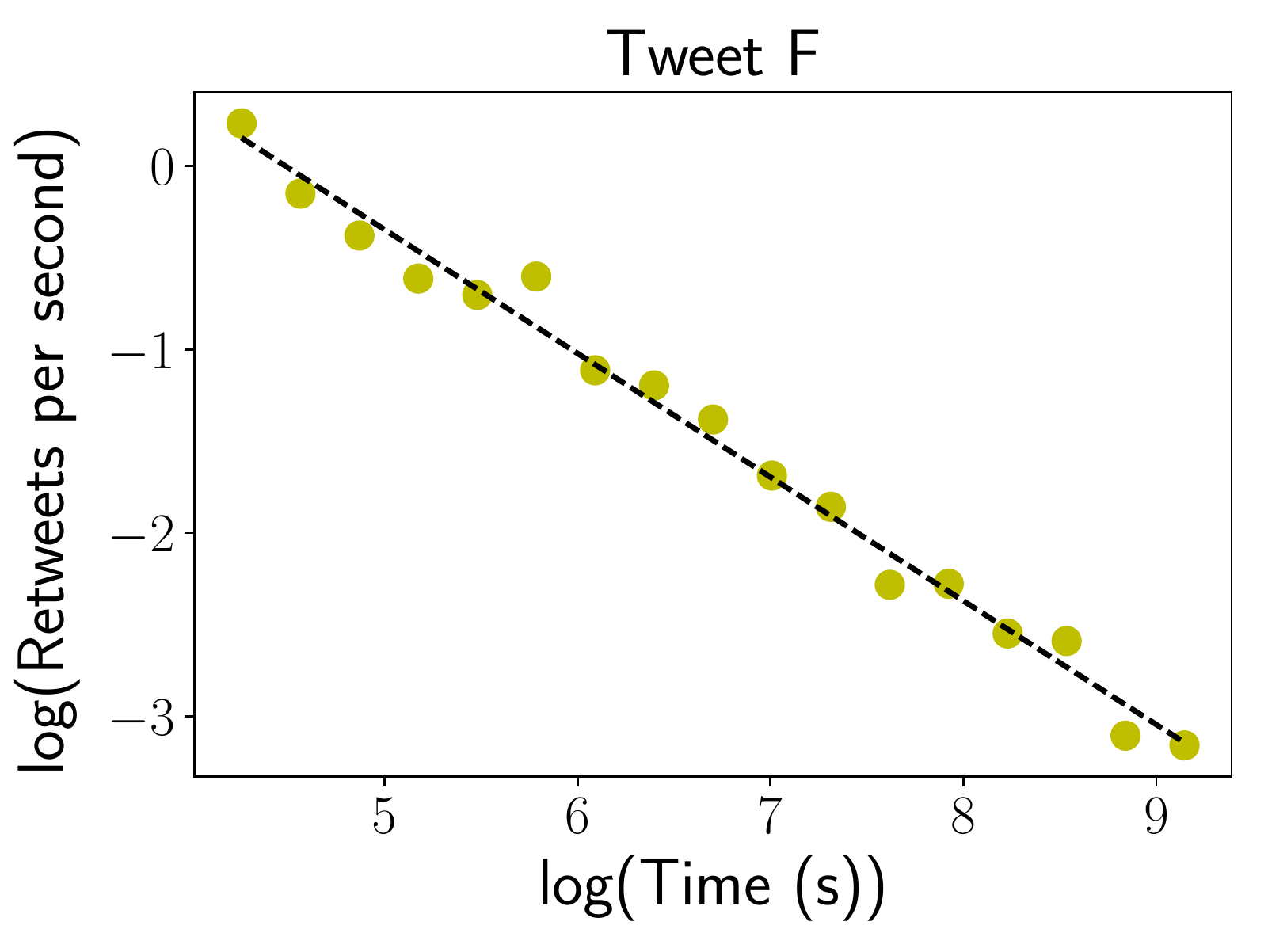} \ \\\
		\end{tabular}
	\end{center}
	\caption{Retweet rate log-log plots over the first 3 hours since the initial tweet. The linear relationship suggests a power law holds within this region.}
	\label{fig:RetweetRateDecay}
\end{figure*}

\begin{figure*}[h]
	\begin{center}
		\begin{tabular}{@{}c@{}c@{}c@{}c}
			\includegraphics[width=2in]{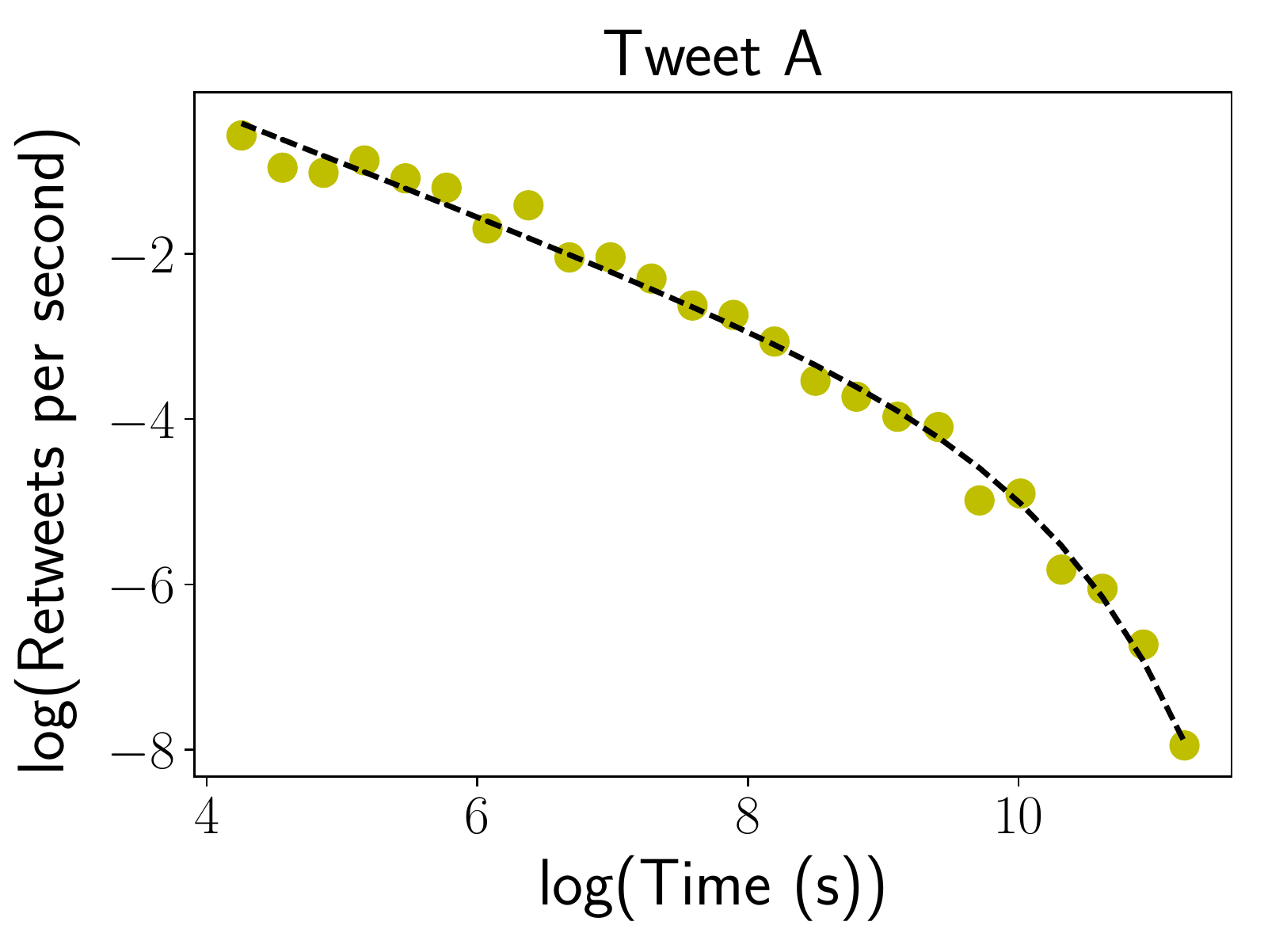} \ &
			\includegraphics[width=2in]{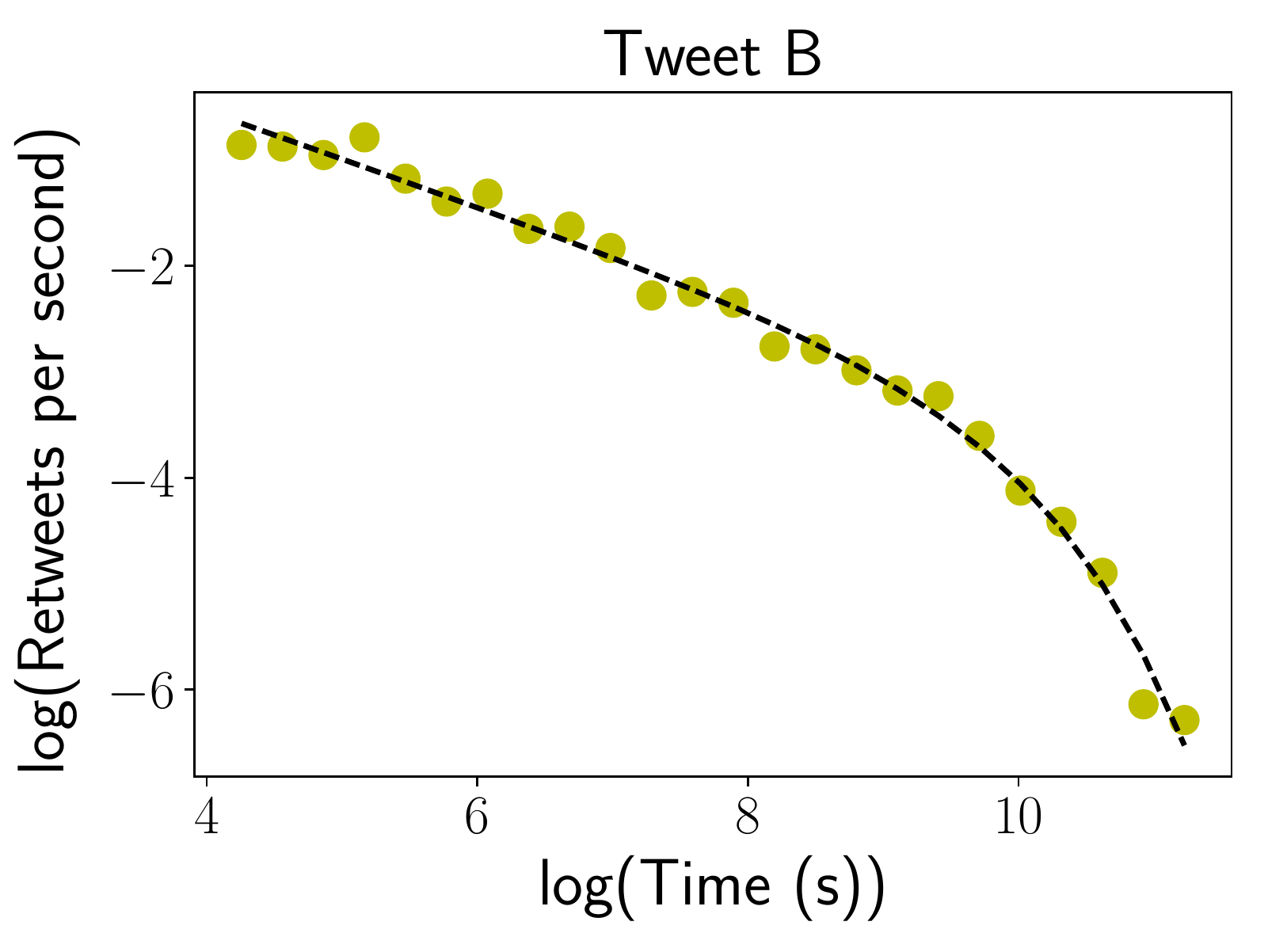} \ &
			\includegraphics[width=2in]{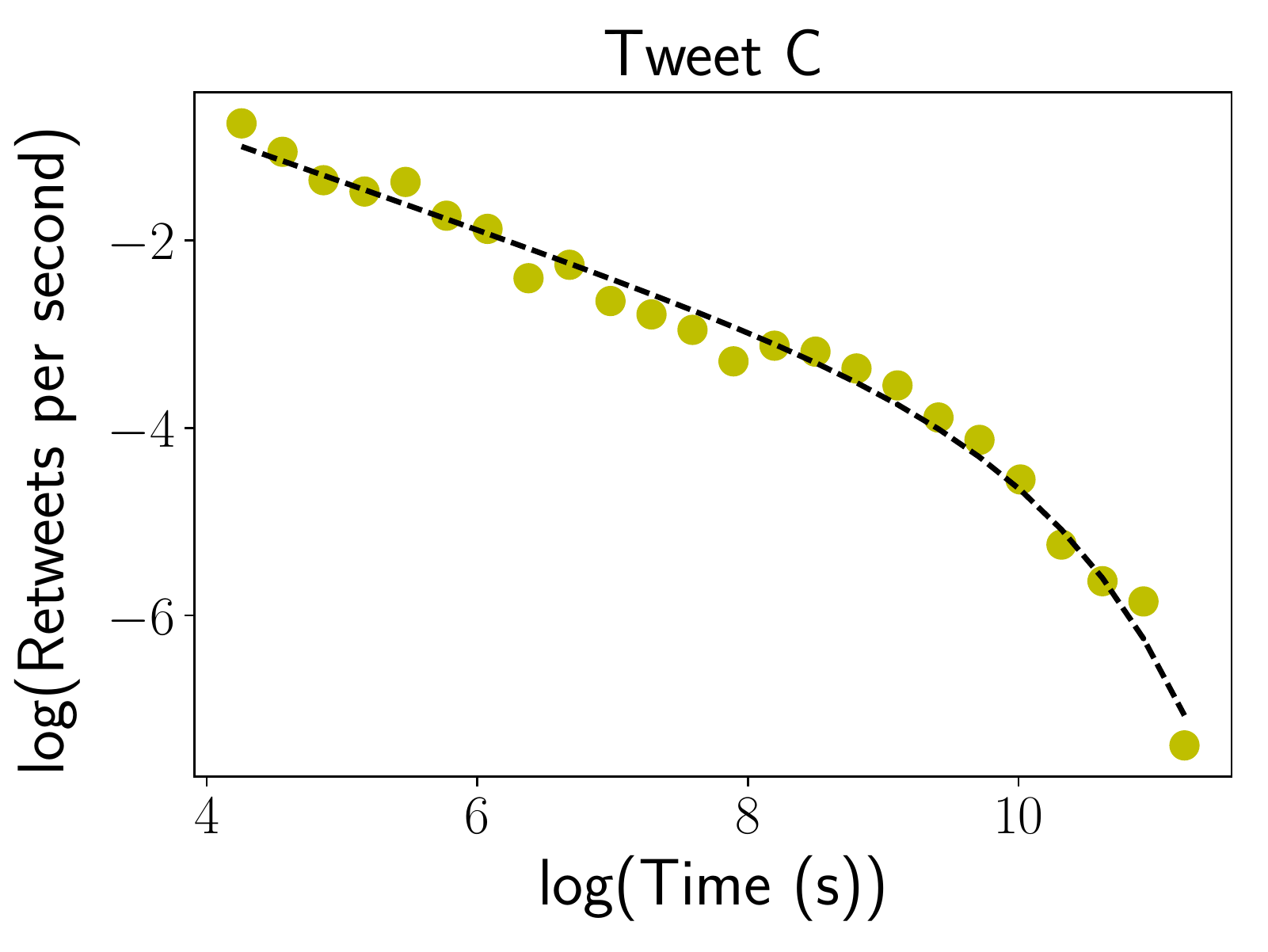} \ \\		
			\includegraphics[width=2in]{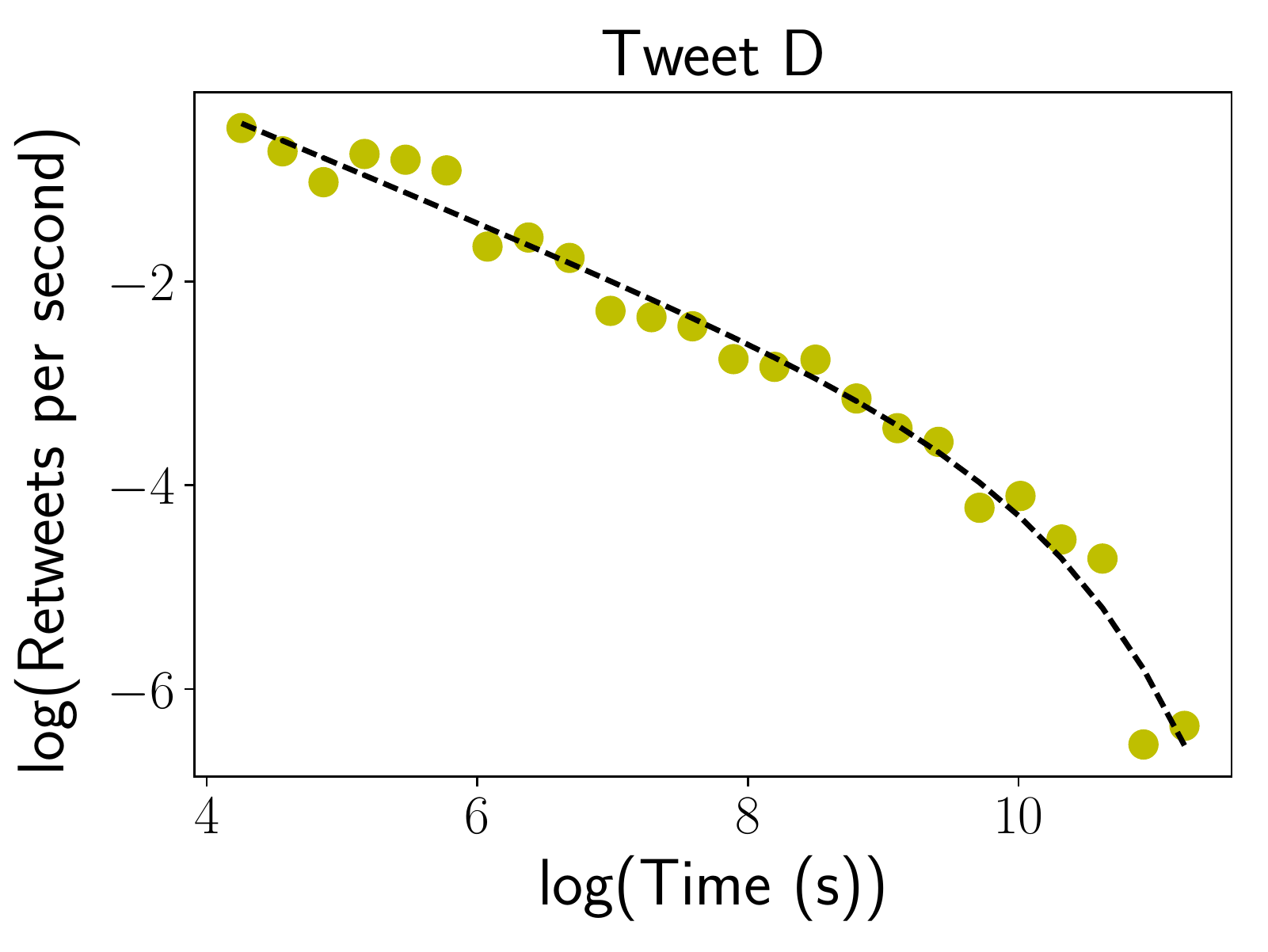} \ &
			\includegraphics[width=2in]{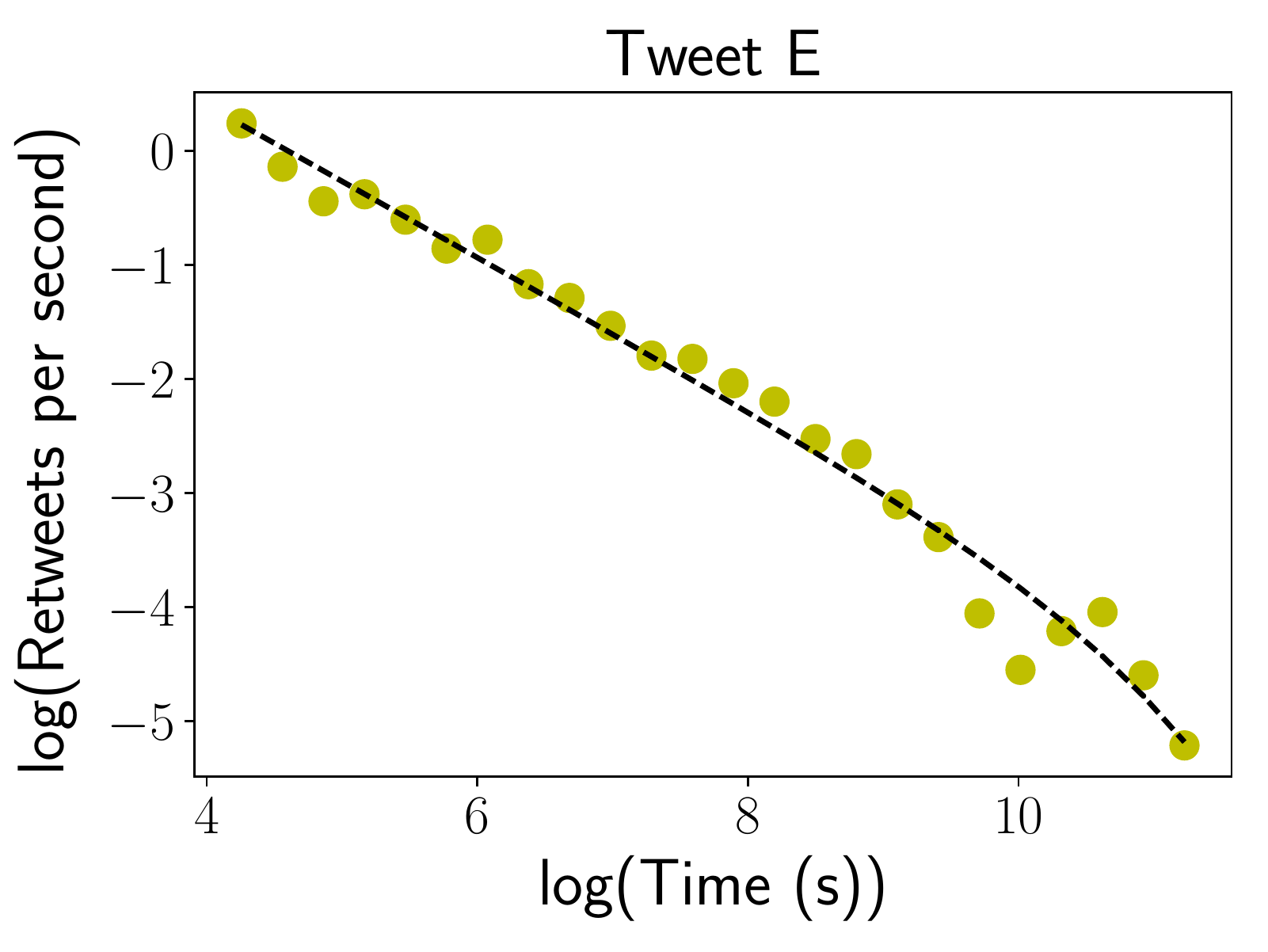} \ &		
			\includegraphics[width=2in]{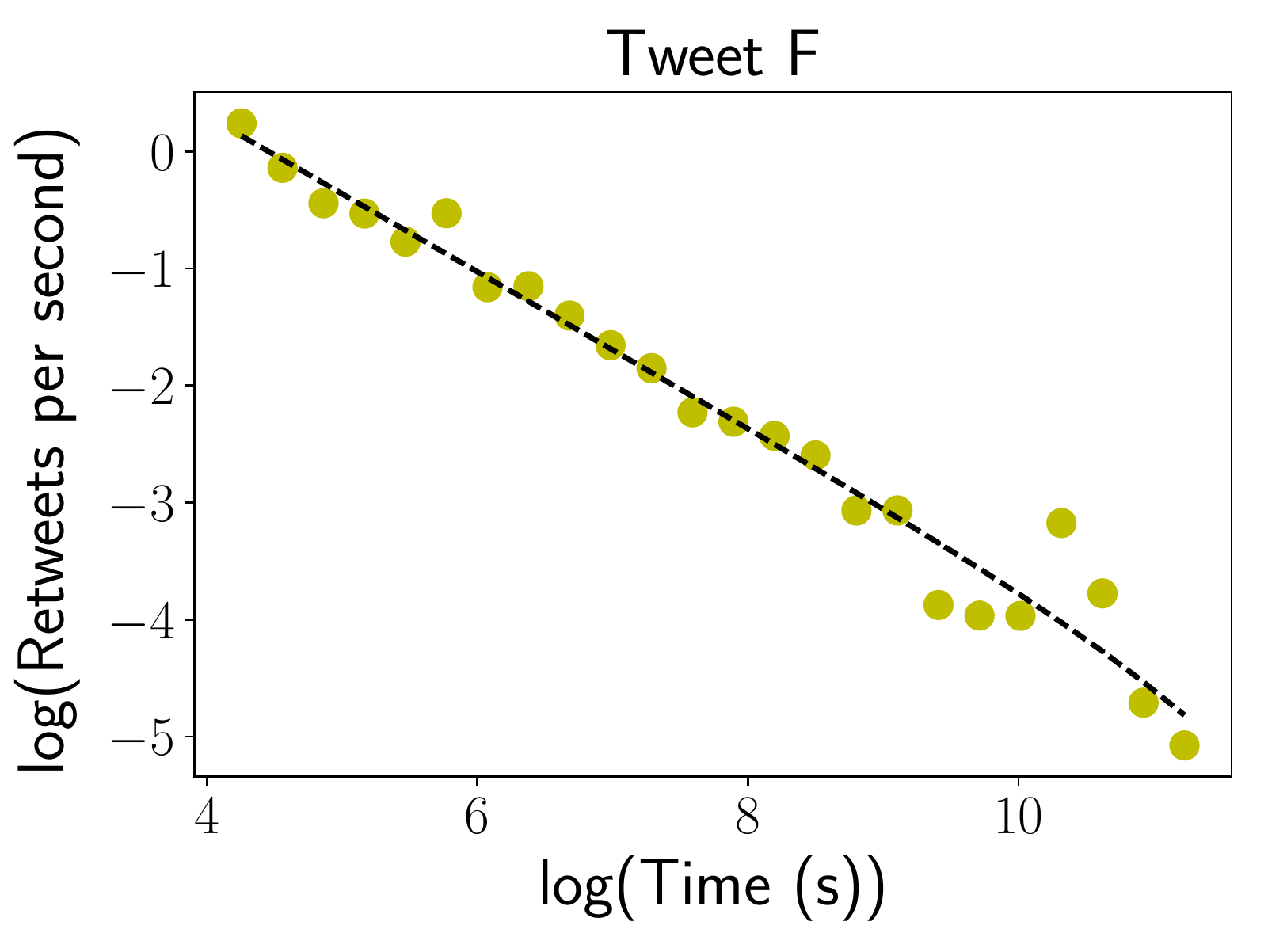} \ \\\
		\end{tabular}
	\end{center}
	\caption{Retweet rate log-log plots over the first 24 hours since the initial tweet. A curve of a power law with exponential cutoff is fitted, showing the faster than linear decay.}
	\label{fig:PowerLawWithCutoff}
\end{figure*}

To illustrate why this distribution might be considered a power law, we choose constant-width bins on a log scale and again plot the log of the retweet rate against the log of time. This gives Figure \ref{fig:RetweetRateDecay}. As can be seen, the graphs are roughly linear, suggesting that in this region the retweet rate is well modelled by a power law.

If we look at the same data set over a longer period, up to 24 hours, and plot the retweet rate on a log-log plot, we get Figure \ref{fig:PowerLawWithCutoff}. This shows visually that the line is no longer straight, so a power law does not appear to continue to fit the data. As we shall demonstrate in Section \ref{sec:powerlawwithcutoff}, this phenomenon tends to occur for the vast majority of seed tweets. As is shown on the graph, a power law with exponential cutoff is a better fit to the data.

In the rest of this section we quantify these claims and show that the power law with exponential cutoff is indeed a better fit than the power law.

\subsection{Collection methodology}

We monitored and collected tweets from the 100 Twitter users with the most followers \cite{TwitterCounter} using the Twitter REST API. We chose these Twitter users as their tweets are retweeted more frequently, providing more dense data. In total, we obtained the times of retweets from a total of 1676 seed tweets in April 2016. We exclude any tweet that was deleted shortly after being tweeted, as this causes a truncated data set. We also exclude any tweet that has less than 100 retweets as it is less meaningful to fit a curve to a sparse data set.

The Twitter REST API allows us to query the details of the 100 most recent retweets from a given tweet. Twitter imposes a rate limit of 15 such queries per 15 minutes, allowing an average of one hundred retweets to be collected per minute. In order to avoid hitting this rate limit, we stop the collection of any retweet set that has a retweet rate greater than 60 retweets per minute, an average of one per second. All remaining retweet times form our dataset to be analysed. This collection methodology gives us the complete retweet cascade for all tweets in our dataset. 

After removing the data which did not fit our criteria we are left with 808 seed tweets, which had a mean of 307.7 retweets and a median of 197 retweets. There were 34 seed tweets with over 1000 retweets. 

We note that as our dataset is only from the subsection of the Twitter population with a high number of followers, we can only make conclusions about information propagation from these users.


\subsection{Fitting a power law} 
\label{sec:powerlaw}

We fit a power law to each of our 808 retweet data sets using maximum likelihood estimation. We choose maximum likelihood estimation to conduct the fit as it is more accurate than logarithmic binning \cite{refId0}. A power law has density function
\begin{equation}
p(x) = Cx^{-\alpha},
\end{equation}
where $\alpha>0$ and $C>0$ is a normalising constant which depends on $\alpha$.

We calculate the Kolmogorov-Smirnov statistic to determine how accurately our empirical distribution matches the theoretical distribution. For a theoretical distribution $F(x)$ and an empirical CDF $S(x)$, the Kolmogorov-Smirnov statistic $D$ is defined by
\begin{equation}
D = \sup_x | F(x) - S(x) |.
\end{equation}

\begin{figure}[h]
	\centering
	\includegraphics[width=3in]{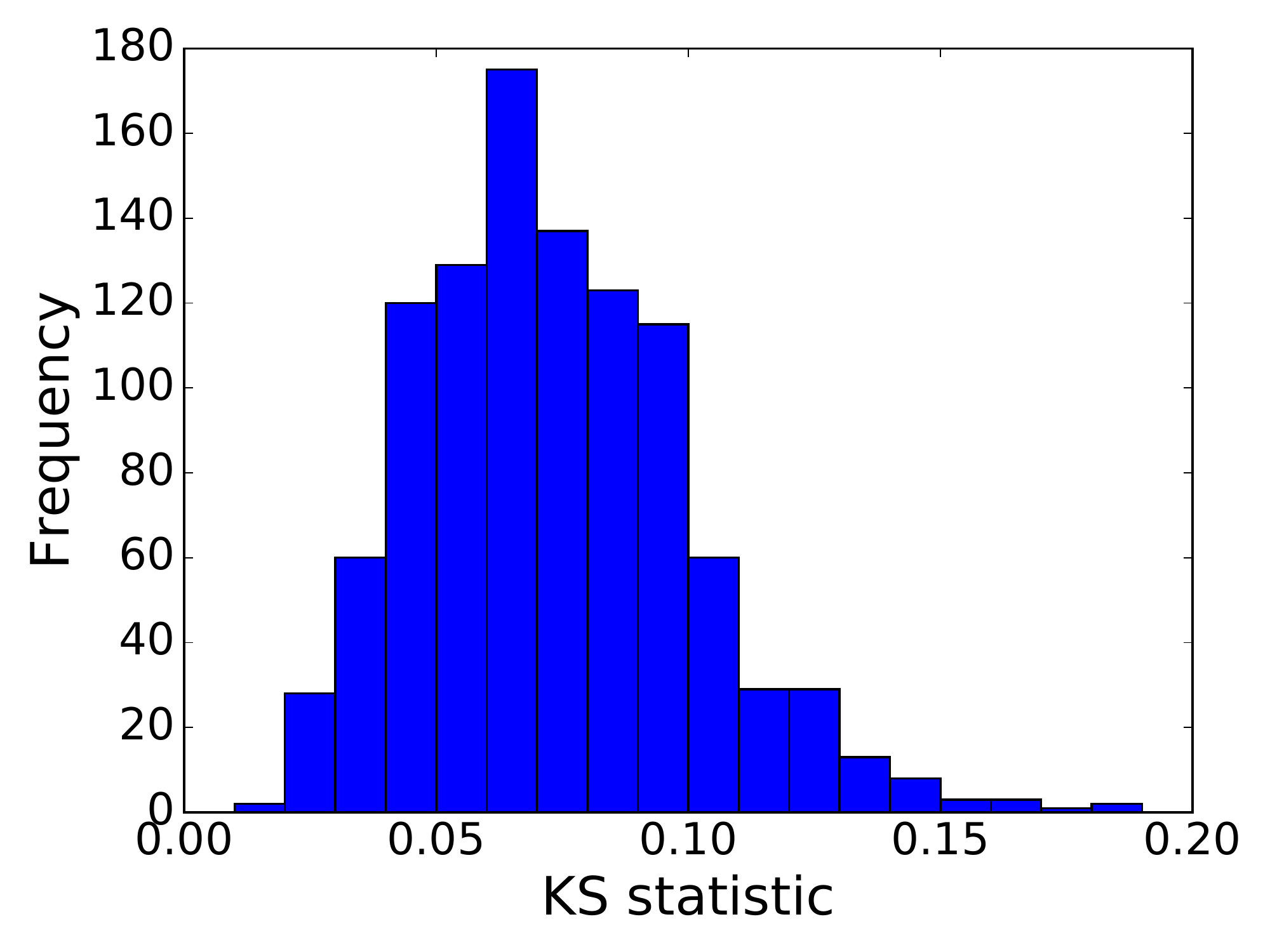}
	\caption{Histogram of KS statistics for a power law fit to the distribution of retweet times. The mean KS value is 0.07454 with standard deviation 0.02966.}
	\label{fig:PowerLawKSHist}
\end{figure}

The histogram of the KS-statistic for each dataset is shown in Figure \ref{fig:PowerLawKSHist}. The mean KS value is 0.07454 with standard deviation 0.02966. As can be seen, the KS-statistic values are centered around this mean and mostly fall between 0.05 and 0.10.

\subsection{Fitting a power law with exponential cutoff}
\label{sec:powerlawwithcutoff}

We also fit a power law with exponential cutoff to each of our 808 retweet time data sets using maximum likelihood estimation. A power law with exponential cutoff has the density function
\begin{equation}
p(x) = Ax^{-b}e^{-cx}.
\end{equation}
with $A,b,c > 0$ and where $A$ is a normalising constant.

\begin{figure}[h]
	\centering
	\includegraphics[width=3in]{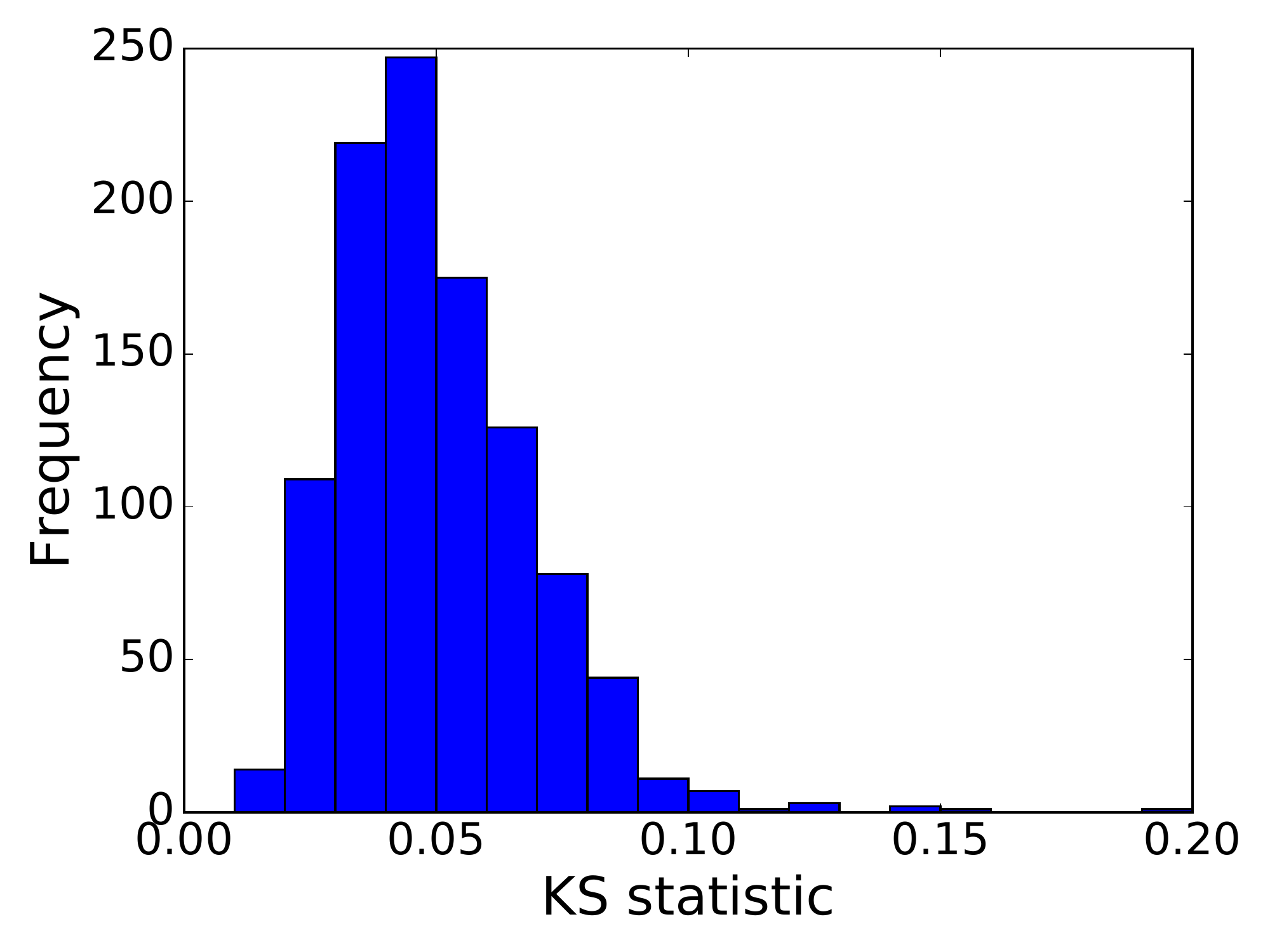}
	\caption{Histogram of KS statistics for a power law with exponential cutoff fit to the distribution of retweet times. The mean KS statistic is 0.05080 with standard deviation 0.02302}
	\label{fig:PowerLawCutKSHist}
\end{figure}

We calculate the KS statistic for each retweet data set modeled by a power law with exponential cutoff. A histogram of the resultant values is shown in Figure \ref{fig:PowerLawCutKSHist}. 

The mean KS statistic is 0.05080 with standard deviation 0.02302. This is lower than the mean KS value of 0.07454 without the exponential cutoff (32\% improvement) and demonstrates a clear improvement in the quality of fit. 

In order to determine whether the reduction in the mean KS-statistic for the power law with exponential cutoff is statistically significant, we conduct a paired t-test on the two sets of data, giving a p-value of $2.26071\times 10^{-157}$. We therefore reject the null hypothesis that the paired differences have zero mean and conclude that the power law with exponential cutoff has a lower KS statistic.

The set of power law distributions is a special case of the set of power laws with exponential cutoffs. As we have added an extra parameter to our model, the power law with exponential cutoff will always provide at least as good a fit. To measure the relative quality of each model, we thus use the AIC criterion
\begin{equation}
\mbox{AIC} = 2k - 2 \ln (L)
\end{equation}
where $k$ is the number of parameters and $L$ is the likelihood function. 

\begin{figure}[h]
	\centering
	\includegraphics[width=3in]{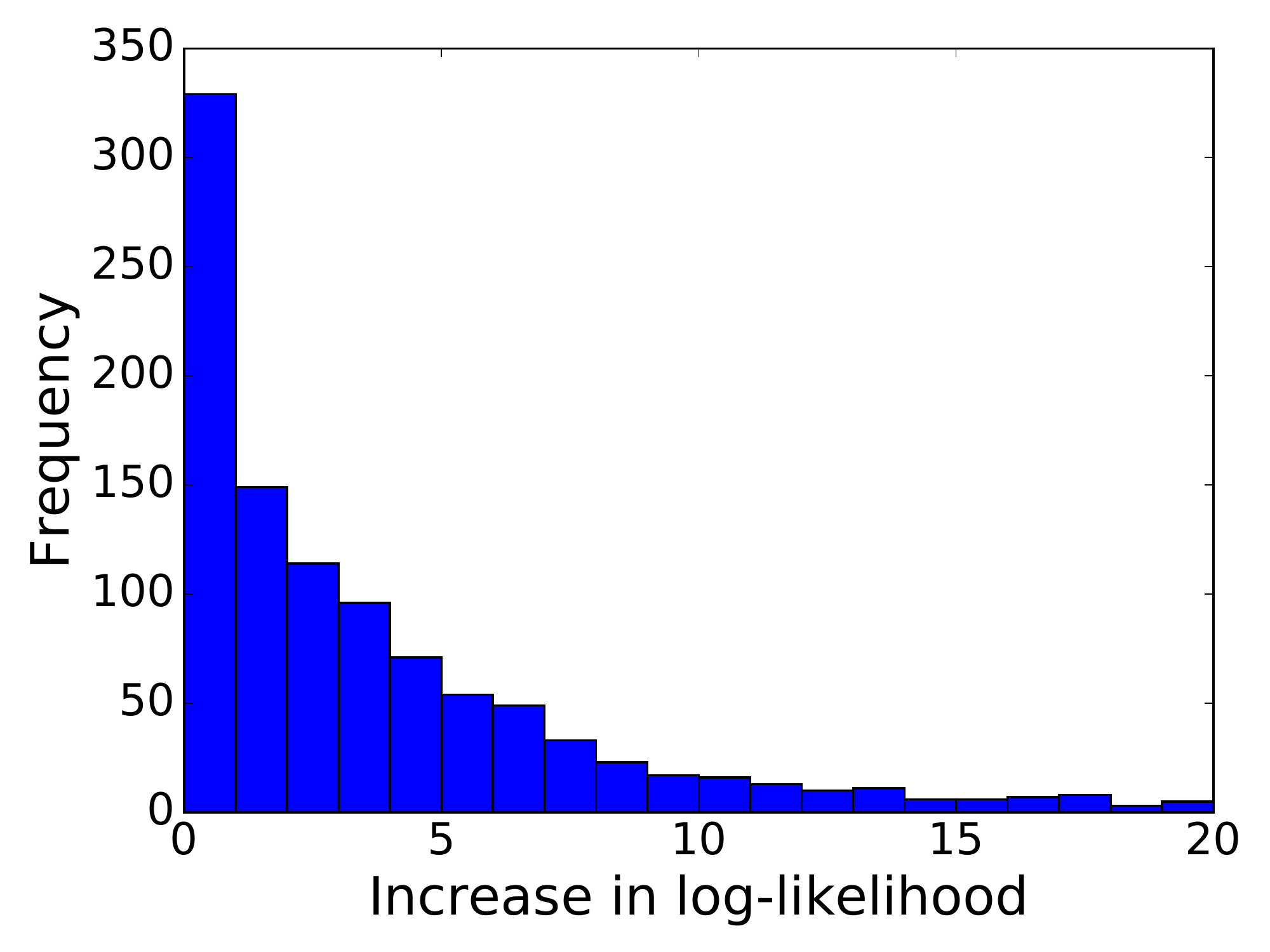}
	\caption{Histogram of improvement to log-likelihood changing from power law to power law with exponential cutoff. The change of distribution improves the likelihood score by more than 1 in 558 out of 808 tested datasets, 69.1\% of cases. The increase in log-likelihood score justifies the additional parameter of the power law with exponential cutoff.}
	\label{fig:CutoffHist}
\end{figure}

We wish to minimise the AIC value. In order to do this, adding an additional parameter requires an improvement in log-likelihood score of $1$ to increase the AIC score. We consider the log-likelihood scores for the power law and power law with exponential cutoff and observe the increase in log-likelihood score in Figure \ref{fig:CutoffHist}.

Some datasets are well modeled by a power law and only show a very small increase in log-likelihood score, while other datasets benefit significantly by adding the cutoff.

Changing from a power law to a power law with exponential cutoff improves the likelihood score by more than 1 in 558 of 808 tested datasets, 69.1\% of cases. It improves the likelihood score by a mean value of 4.239. Consequently, adding an exponential cutoff improves the AIC score by a mean value of 6.478.

We conclude that adding an exponential cutoff to the power law provides a better fit.

\section{Explanation of power law with exponential cutoff}
\label{sec:originofpowerlaws}
\label{prioritytasking}

A potential cause of the power law in retweet activity is due to a decision-based queuing process. The action of checking Twitter and deciding whether to retweet is a task prioritised against other daily activities. Consequently the time between a tweet arriving and a user checking their twitter account has a power law distribution \cite{barabasi05}.

A decision-based queuing process is much more relevant to describe human activity on the internet than the more commonly discussed cause of power laws, preferential attachment \cite{10.1371/journal.pone.0064349, 1999PhyA..272..173B}. Users will implicitly assign priorities to tasks in their lives and execute these tasks according to their internal perceived priorities. This explains the origin of the power law component for the distribution of time until retweets. 

The second factor affecting the retweet distribution is the loss of interest in topics over time, which has exponential decay \cite{Ding:2005:TWC:1099554.1099689, Li:2014:MBT:2566267.2566312}. If the topic of the tweet is less relevant than when it was tweeted, it is less likely that it will be retweeted. The third and final component that affects the likelihood of a retweet is the proportion of users who decide to retweet. For our explanatory model, we assume that a constant proportion of users who see the tweet at a time when it is still relevant will decide to retweet. 

To obtain the overall likelihood of retweet at time $t$, we multiply these three components together:

\begin{equation}
\begin{split}
P(\text{Retweet at time $t$}) = P(\text{Twitter checked at time $t$}) \\
\times P(\text{Tweet still relevant at time $t$}) \\
\times P(\text{User will choose to retweet}).
\end{split}
\end{equation}

This gives
\begin{equation}
P(\text{Retweet at time $t$}) = At^{-b}e^{-ct}.
\end{equation}

It is possible that there are alternative explanations for the cause of the power law with exponential cutoff. However, our explanation is simple and explains every component of the phenomenon that we have observed in the empirical data. 

\section{Discussion and Conclusions}
\label{sec:conclusions}

The rate of retweets can be well modelled by a power law with exponential cutoff, providing a better fit than a standard power law distribution. The power law component is explained by the time until the user checks their social media, which is governed by a decision-based queuing process. The exponential cutoff is explained by the loss of interest in topics over time.

In this work we analysed retweet times from the 100 Twitter users with the most followers. A natural question is whether similar retweet rate distributions would hold for all other Twitter users. 

Future work will analyse how the parameters of the power law and exponential cutoff vary based on author, tweet topic or other factors. This will allow prediction of the propagation rate of the tweet. We could also look at population-level social questions, e.g. how do the decay parameters vary over the long term? As a society, are we growing more or less engaged with news from social media? As the tweet/retweet mechanism provides a continual source of information propagation data, it is possible to test theories which have been proposed in the social science literature using this experimental environment.

The model that we have produced gives an explanation of the phenomena that govern the spread rate of information online through Twitter. It builds upon previous work on the burstiness of human behaviour to give a better understanding of cascades in a social media information system.

\section{Acknowledgments}
PM, LM, and NGB acknowledge the financial support of the Data to Decisions Cooperative Research Centre (D2DCRC). All the authors acknowledge the financial support of the ARC Center of Excellence for Mathematical and Statistical Frontiers (ACEMS).

\vfill\eject
%

%
\end{document}